\documentclass{ws-procs975x65}
\usepackage{graphicx}

\usepackage{amssymb}
\usepackage{simbolos09}

\begin{document}

\title{CURVATURE(S) OF A LIGHT WAVEFRONT IN A WEAK GRAVITATIONAL FIELD}

\author{J.-F. PASCUAL-S\'ANCHEZ, A. SAN MIGUEL and F. VICENTE }

\address{Dept. de Matem\'atica Aplicada, Universidad de Valladolid,\\
 47005 Valladolid, Spain\\
E-mails: asmiguel@maf.uva.es, fvicente@maf.uva.es, jfpascua@maf.uva.es}


\begin{abstract}
The geometry of a light wavefront evolving from a flat wavefront under the action of weak gravity field in the 3-space associated to a post-Newtonian relativistic spacetime,  is studied numerically by means of the ray tracing method.

\end{abstract}

\keywords{Ray tracing; Light wavefront; Post-Newtonian formalism; Numerical Relativity.}

\bodymatter

\section{Introduction}\label{aba:sec1}
The curvature of initially plane light wavefronts by a gravity field is a purely general relativistic effect that has no special relativistic analogue.
In order to obtain an experimental measurement of the curvature of a light wavefront, Samuel recently
proposed a method  based on the relation between the differences of arrival time recorded at four points on the Earth and the volume of a parallelepiped determined by four points in the curved wavefront surface, see Ref.~\citen{Sam}. In this work and in   Ref.~\citen{san} with more detail, we study a discretized model of the wavefront surface by means of a regular triangulation for the study of the curvature(s) (mean and relative, see Ref.~\citen{dC})  of this surface.

\section{Light propagation in a weak gravitational field}

Let us consider a spacetime $(\mathcal{M},g)$ corresponding to a weak gravitational field determined by a metric tensor given in a global coordinate system  $\{(\bm{z},ct)\}$ by
                $g_{\alpha\beta}=\eta_{\alpha\beta}+h_{\alpha\beta}$,  with $\eta_{\alpha\beta}=\diag(1,1,1,-1)$.
            Where  the coordinate components of the metric perturbation $h_{\alpha\beta}$ are:
        \begin{equation}
            h_{ab} =  2c^{-2}\kappa \|\bm{z}\|^{-1}\delta_{ab},\quad
            h_{a4} = -4c^{-3}\kappa \|\bm{z}\|^{-1}\dot{Z}_a, \quad
            h_{44} =  2c^{-2}\kappa \|\bm{z}\|^{-1},
        \end{equation}
        here $\kappa:=GM$ represents the gravitational constant of the Sun, located at $Z^a(t)$,
         and \,$c$\, represents the vacuum light speed.
The null geodesics, $z(t)=\big(\bm{z}(t),t\big)$ satisfy the following equations, see Ref.~\citen{Bru}:
    \begin{eqnarray*}
          \ddot{\bm{z}}^a & =& \mfrac{1}{2} c^2h_{44,a} -[\mfrac{1}{2}h_{44,t}\delta^a_k+h_{ak,t}+c(h_{4a,k}-h_{4k,a})]\dot{\bm{z}}^k\nonumber\\
        & & -(h_{44,k}\delta^a_l+h_{ak,l}-\mfrac{1}{2}h_{kl,a})\dot{\bm{z}}^k\dot{\bm{z}}^l \nonumber\\
        & & -(c^{-1}h_{4k,j}-\mfrac{1}{2}c^{-2}h_{jk,t})\dot{\bm{z}}^j\dot{\bm{z}}^k\dot{\bm{z}}^a,\label{PN1a}\\[.1in]
0 & = &  g_{\alpha\beta}\dot{z}^\alpha\dot{z}^\beta.\label{PN1b}
\end{eqnarray*}
    The second equation is the isotropy constraint satisfied by the null geodesics.

\section{Local approximation of the wavefront}
Let be $\mathcal{S}_0$ a flat initial surface far from the Sun, formed by points $(z_1,z_2,-\zeta)$ (with $\zeta>0$) in an asymptotically Cartesian coordinate system $\{z\}$. For the discretization of $\mathcal{S}_0$ a triangulation is constructed in such a form that each vertex is represented by a complex number of the  set:
\begin{equation}\label{vertices1}
\bar{\mathcal{V}}:=\{z=a_1+a_2\omega+a_3\omega^2\;|\;\; a_1,a_2,a_3\in\mathcal{A},\; \omega:=\exp(2\pi \uniC/3)\},
\end{equation}
The initial triangulation by $\mathcal{V}$ induces a triangulation on the final wavefront $\mathcal{S}_t$.
The evolution of  a photon  $\bm{z}_0:=\bm{z}(0)\in\mathcal{V}$ with velocity $\dot{\bm{z}}_0:=(0,0,c)$ in phase space $\bm{u}=(\bm{z},\dot{\bm{z}})$ may be written as a {first order} differential system {$\dot{\bm{u}}=\bm{F}(\bm{u},t)$}. This determines a flow, $\bm{z}(t) =\varphi_t(\bm{z}_0,\dot{\bm{z}}_0)$, in the 3-dimensional curved quotient space of $ \mathcal{M} $ by the global timelike vector field  $ \partial_t$ associated to the global coordinate system used in the post-Newtonian formalism.
For each time $t$, the flow $\varphi_t$ determines a 2-dimensional curved wavefront $\mathcal{S}_t$.

To compute the curvatures of the wavefront surface corresponding to the mesh $\mathcal{V}$
at each inner vertex, we consider a 1--ring  formed by the six vertices  closest. For each 1--ring,  one obtains on the mesh $\mathcal{V}$ the image under the flow $\varphi_t$.
In a neighbourhood of the image point  the wavefront can be approximated by a least-squares fitting of the data obtained as the quadric:
        \begin{equation}\label{2}
            y^3=f(y^1,y^2):=\mfrac{1}{2}a_1(y^1)^2 + a_2y^1 y^2 + \mfrac{1}{2}a_3(y^2)^2.
        \end{equation}
        using adapted normal coordinates $\{y^i\}$.

\section{Numerical integrator}
We apply the ray tracing method, see Refs.~\citen{JZ,san},  to a  tubular region of light wavefront region supposing a gravitational model generated by a static Sun, considered as a point.
The mean and relative total curvatures, defined in Refs.~\citen{san,dC}, are computed  at each inner vertex of mesh on the light wavefront surface $\mathcal {S}_t$ in the vicinity of the Sun, by the implementation of the following  pseudocode:

\begin{center}
    \begin{tabular}{ll}
    \hline %
     {\bf Data: } $\bm{u}^*_n:=(\bm{z}^*_n,\dot{\bm{z}}^*_n), n=1,\dots N$          & \\
     {\bf for } $n=1\dots N$ {\bf do}                                               & \\
     \qquad $\bm{u}_n :=\rm{\tt Taylor}(t,\bm{u}_n^*)$                            & \\
     \qquad $\bm{y}_n := \rm{\tt NormalCoordinates}(\bm{z}_n)$                    & \\
     \qquad {\bf for } $i=0\dots 6$ {\bf do}                                        & \\
     \qquad \qquad $\bm{y}_{n_i} := \rm{\tt Ring}(\bm{y}_n)$                      & \\
     \qquad {\bf end}                                                               & \\
     \qquad $(a_1,a_2,a_3) = {\tt LeastSquares}(\bm{y}_{n_i})$                      & \\
     \qquad $\gamma_{AB}(\bm{x}_n) := \rm{\tt Metric}(a_1,a_2,a_3,\bm{x}_n)$      & \\
     \qquad $B= {\tt SecondFundamentalForm}(\bm{x}_n)$                               & \\
     \qquad $(\lambda_1,\lambda_2) = {\tt Diagonalize}(B)$                          & \\
     \qquad $(K_{\rm{rel}},H) = {\tt Curvature}(\lambda_1,\lambda_2)$             & \\
     {\bf end}                                                                      & \\
     \hline
\end{tabular}
\end{center}

        \begin{figure}[h]
\begin{center}
\psfig{file=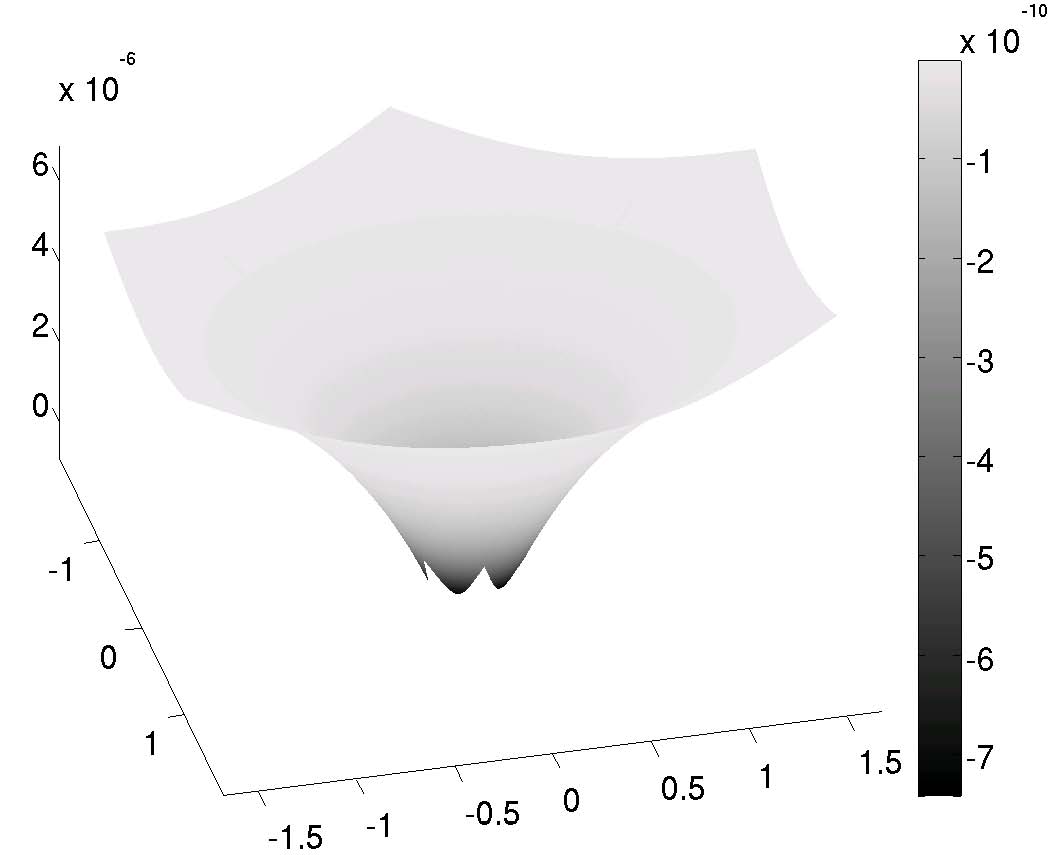,width=2.5in}
\caption{Wavefront surface and relative curvature (gray scale) deformed by a spherical gravitational field (a different scale is used for the vertical axis).}\label{figure4}
\end{center}
\end{figure}

In Figure~\ref{figure4} the surface $\mathcal{S}_T$ at the time when the wavefront arrives at the Earth is shown using a gray-scale to represent the relative curvature (note we have used a different scale on the $Oz^3$--axis). One sees in this figure that the absolute value of the relative curvature  defined on $\mathcal{S}_T$ increases as the distance between the photon and the $Oz^3$--axis, where the Sun is located, decreases.

\section*{Acknowledgements}
This research was partially supported by the Spanish Ministry de Educaci\'on y Ciencia, MEC-FEDER grant ESP2006-01263.

\end{document}